\documentclass[twocolumn,final]{article}
\usepackage{preprint}
\usepackage{wrapfig}
\usepackage{graphicx}
\usepackage[hidelinks]{hyperref}

\usepackage{ulem}
\usepackage{tabularray}
\usepackage{multirow}
\usepackage[bottom]{footmisc}
\usepackage{cite}
\usepackage{amsmath,amssymb,amsfonts}
\usepackage{cleveref}
\usepackage{algorithmic}
\usepackage{textcomp}
\usepackage{flushend}
\usepackage{xcolor}

\def\BibTeX{{\rm B\kern-.05em{\sc i\kern-.025em b}\kern-.08em
    T\kern-.1667em\lower.7ex\hbox{E}\kern-.125emX}}
\newcommand{\name}{\textsc{Oxn}}
\DeclareTextFontCommand{\models}{\ttfamily\hyphenchar\font=45\relax}

\title{Informed and Assessable Observability Design Decisions in Cloud-native Microservice Applications}

\usepackage{authblk}

\author[1]{Maria C. Borges}
\author[1]{Joshua Bauer}
\author[1]{Sebastian Werner} 
\author[1]{Michael Gebauer}
\author[1]{Stefan Tai}
\affil[1]{Information Systems Engineering,
Technische Universität Berlin, Germany}

\begin{document}
\twocolumn[\begin{@twocolumnfalse}
\maketitle
\pagestyle{plain}

\begin{abstract}
Observability is important to ensure the reliability of microservice applications. These applications are often prone to failures, since they have many independent services deployed on heterogeneous environments. When employed “correctly“, observability can help developers identify and troubleshoot faults quickly. However, %
instrumenting and configuring the observability of a microservice application is not trivial but tool-dependent and tied to costs. 
Architects need to understand observability-related trade-offs in order to weigh between different observability design alternatives. 
Still, these architectural design decisions are not supported by systematic methods and typically just rely on ``professional intuition".  

In this paper, we argue for a systematic method to arrive at informed and continuously assessable observability design decisions. Specifically, we focus on fault observability of cloud-native microservice applications, and turn this into a testable and quantifiable property. %
Towards our goal, we first model the scale and scope of observability design decisions across the cloud-native stack. Then, we propose observability metrics which can be determined for any microservice application through so-called observability experiments. We present a proof-of-concept implementation of our experiment tool \name{}. \name{} is able to inject arbitrary faults into an application, similar to Chaos Engineering, but also possesses the unique capability to modify the observability configuration, allowing for the assessment of design decisions that were previously left unexplored. %
We demonstrate our approach using a popular open source microservice application and show the trade-offs involved in different observability design decisions.

\end{abstract}
\vspace{0.5cm}

\end{@twocolumnfalse}]

\section{Introduction}The observability practices of monitoring, tracing and logging play an important role in ensuring the reliability of cloud-native microservice architectures. 
Microservices provide many advantages over their monolithic predecessors~\cite{Fowler_Microservices_2014}, however the resulting applications can be difficult to troubleshoot when a fault occurs\cite{Niedermaier_ObservabilityInterviewStudy_2019,Zhang_MicroserviceSurvey_ThresholdingHard_2019}. This is because each request often travels through multiple independently developed services, deployed on an intricate and evolving stack of software platforms. To address this challenge, observability systems have evolved accordingly, %
now consisting of distributed services that collect observability data from all levels of runtime platforms as well as from the application level through built-in and developer-defined instrumentation.
Hence, using these observability systems requires a multitude of hard decisions in terms of instrumentation points, observability services and configuration. 
When employed correctly, observability can help developers identify faults quickly, but improper settings can also obfuscate faults \cite{Zhang_MicroserviceSurvey_ThresholdingHard_2019,Borges_TracingServerless_2021} and increase latencies and cost.

Still, designing %
the observability of microservice applications is not trivial. It implies often overlooked tasks like the tool-dependent configuration of parameters, setting alerts and adding custom instrumentation code. These decisions need to be weighed against observability-related trade-offs, e.g., the performance overhead on the application, plus the cost overhead associated with operating the observability infrastructure.   %
However, these architectural design decisions are not supported by systematic methods. Instead, decisions are made based on previous experience and “professional intuition” of the practitioners \cite{Zhang_MicroserviceSurvey_ThresholdingHard_2019,Vale_MicroserviceQualityMeasurementIndustrySurvey_2022}, or developers may end up eagerly instrumenting and changing configuration parameters in a reactive manner after a problem occurs \cite{Niedermaier_ObservabilityInterviewStudy_2019}. To weigh between different design alternatives and arrive at an appropriate configuration that justifies the effort, practitioners must have a method to assess the effectiveness of their observability. 

Consequently, the need to evaluate the degree of observability of a system has been recognized by research and industry alike \cite{Tamburri_MVCObservability_2018,Google_MonitoringMeasurement_2023,Ahmed_EffectivenessOfAPM_2016,Vale_MicroserviceQualityMeasurementIndustrySurvey_2022}. With suitable metrics, it would be possible to obtain a concrete understanding of the quality of the observability, which could then be tracked and continuously improved over time. However, observability has been challenging to quantify so far \cite{Tamburri_MVCObservability_2018}. Most research has focused either on qualitative assessments of observability tooling \cite{Janes_TracingQualitativeAnalysis_2022}, or solely looks at cost and overhead  \cite{Ernst_OfflineTraceGeneration_2021,Reichelt_OverheadOpenTelemetryEtc,Dinga_EnergyEfficiencyOfMonitoring_2023}, failing to address the question of \textit{effectiveness} of observability. 
Currently, there are no mechanisms for practitioners to quantify or compare the observability of their applications.

In this paper, we argue for a systematic, reproducible and comparative assessment of observability design decisions. %
Our focus lies on the observability of faults specifically. Similar to software quality measures like test coverage, we aim to make fault observability a testable and quantifiable system property. This, in turn, would help developers identify unobservable faults in their application, thus guiding the configuration and instrumentation process. 

Towards this goal, we present the following contributions: \begin{enumerate}
    \item A model for understanding the scale and scope of observability design decisions.
    \item The concept for testable and quantifiable fault observability metrics.%
    \item Design, implementation and demonstration of \name{}, a tool to automate the process of observability assessments.
\end{enumerate}
The paper is structured as follows: Section \ref{sec:background}  discusses related work.  Section \ref{sec:model1}  provides background and models the observability design space of cloud-native architectures. Section \ref{sec:concept} presents our observability metrics and experiment method. Section \ref{sec:architecture} introduces our supporting experiment tool \name{}. Section \ref{sec:eval} evaluates our method and tool through exemplary experiments. Section \ref{sec:discussion} discusses suitability and limitations. Section \ref{sec:conclusion} concludes with future work.
\label{sec:intro}
\section{Related Work}\label{sec:background}

When faults occur in complex microservice compositions, they can be more complicated to identify than in traditional monoliths, so observability practices have evolved accordingly (e.g., \cite{2010_Sigelman_Dapper,Li_ServiceMesh_2019,Kaldor_Canopy_2017}) and observability remains a growing subject in reliability literature. Multiple design decisions have to be made when implementing and using observability tooling. Different model-driven approaches have been proposed to automate the implementation of observability designs  \cite{Klein_ModelDrivenObservability_2016,Phipathananunth_SyntheticModelBasedMonitroingMicroservices_2018}, yet none of these approaches help practitioners arrive at appropriate and assessable decisions.

Observability design decisions have significant consequences because, as Niedermeier et al. \cite{Niedermaier_ObservabilityInterviewStudy_2019} point out, ``careless deployment and configuration of monitoring agents have been mentioned as potential problems which might cause instability and an increasing network load". So far, different approaches have been proposed to deal with this design challenge.

Haselböck and Weinreich \cite{Haselboeck_DecisionModelMicroservices_2017} propose decision guidance models for microservice observability decisions. These models help developers with the more abstract high-level decisions, e.g., what tool to adopt when the goal is to inspect service interactions. The model serves well as an introductory resource, but it can't provide assistance with more intricate configuration or instrumentation decisions. For instance, it does not offer guidance on selecting appropriate metrics or determining the ideal sampling rate. In a similar vein, \cite{Sambasivan_TracingDesignChoices_2016} provide best practices for supporting tracing design decisions. Besides these guides, developers can also refer to surveys like \cite{Janes_TracingQualitativeAnalysis_2022} for making decisions, where observability tools are compared based on a number of different qualitative criteria. 

Some research also addresses observability design trade-offs, in particular the costs or performance overhead incurred. Ernst and Tai \cite{Ernst_OfflineTraceGeneration_2021} propose an offline approach to tracing overhead assessment. The model-based approach generates realistic trace data, which can then be used as a workload against different tracing backends. A concrete benchmark of observability instrumentation has also been conducted by \cite{Reichelt_OverheadOpenTelemetryEtc}. Here, the researchers propose a tool for continuous measurement of the overhead of popular instrumentation libraries like \textsc{OpenTelemetry} and \textsc{Kieker}. More recently, the energy efficiency of observability tools has also been investigated \cite{Dinga_EnergyEfficiencyOfMonitoring_2023}. The researchers discovered a close association between observability-related energy consumption and performance overhead, an outcome that was anticipated.
All three of these papers address the drawbacks of observability but overlook the advantages it offers. It is not feasible to evaluate trade-offs without understanding the value provided by something.

Our work is the first to guide observability design choices through comparative measurement of its effectiveness. Ahmed et al. \cite{Ahmed_EffectivenessOfAPM_2016} did similar work, as they measure how effective four different monitoring tools are at identifying performance regressions.  However, their work primarily focuses on comparing the tools without delving into the details of instrumentation and configuration decisions necessary for implementing each tool, as only the default configuration and out-of-the-box automatic instrumentation is used for the assessment.
Another promising approach to improving resilience and testing the configuration of the observability systems is Chaos Engineering. It provides a method  \cite{Basiri_ChaosEngineeringPrinciples_2016} and tools \cite{Heorhiadi_GremlinChaosEngineering_2016,Meiklejohn_FillibusterChaosEngineering_2021,Simonsson_ChaosOrca_2021,Zhang_PhoebeChaosEngineering_2022,Zhang_PhoebeChaosEngineering_2022} for carrying out resilience experiments. Here, faults are injected into a running microservice system to test a hypothesis on the systems ability to withstand these faults. 
However, this approach presupposes a sufficient level of observability prior to conducting the experiments and does not offer any guidance on how to achieve this.

Lastly, Nedelkoski et al. \cite{Nedelkoski_MultiSourceSyntheticData_2020} highlight the lack of datasets incorporating telemetry data from diverse sources (metrics, traces, logs). They propose a solution by developing a microservice system that simulates injected faults, enabling the creation of new datasets, similar to our approach. 
Unfortunately, the system's inability to change or configure observability and specific telemetry instrumentation points greatly limits its ability to assess observability effectiveness.

\section{Modeling Observability Design Decisions}\label{sec:model1}
\begin{figure*}[ht!]
    \centering
    \includegraphics[width=\textwidth]{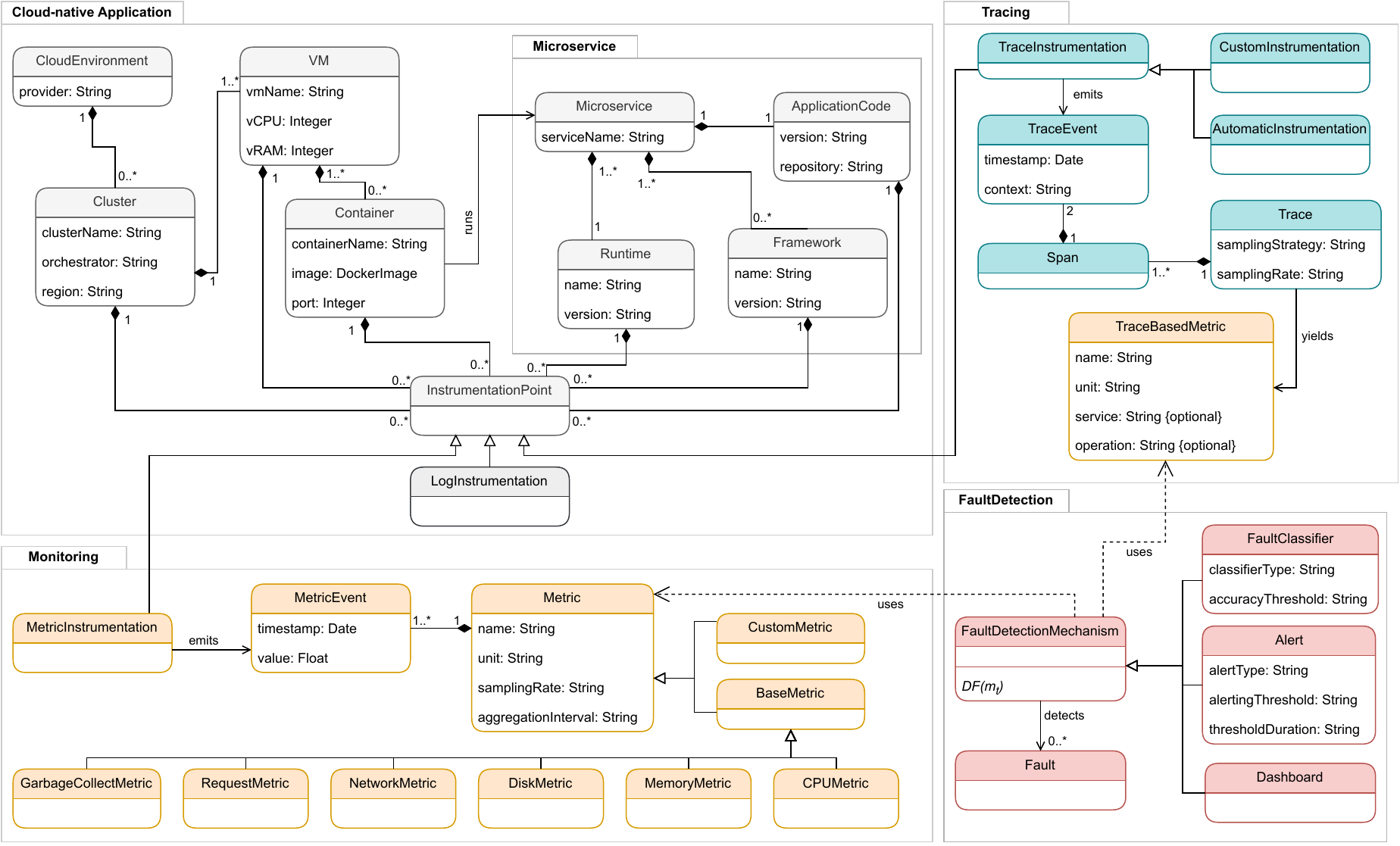}
    \caption{Model of Observability Design Decisions in Cloud-Native Applications}
    \label{fig:model}
\end{figure*}
Ensuring the reliable operation of microservice-based applications is a complicated task that requires observability. Observability, as defined in literature, is the ability to measure the internal state of a system by its outputs \cite{Niedermaier_ObservabilityInterviewStudy_2019}. The process of defining these outputs is what we define as ``observability design decisions". 
In this section, we delve into the intricacies and nuances of these decisions by modeling a typical cloud-native microservice application deployed in accordance with modern practices. 
With this model, we aim to show the scale and scope of observability design decisions, plus it will serve as a common language for practitioners to understand and discuss observability design alternatives. Figure \ref{fig:model} shows our model, which we break down in the following paragraphs.

A \textbf{Cloud-native Application} is deployed in a \models{Cloud Environment} managed by a third-party provider. The model represents the application topology as one or more deployment clusters (\models{Cluster}), deployed in a specific region and managed by an orchestrator. The orchestrator is a deployment technology, e.g. Kubernetes\footnote{\url{https://kubernetes.io}} or Docker Compose\footnote{\url{https://docs.docker.com/compose/}}, that automates the deployment and management of \models{Containers} across virtual machines (\models{VMs}). Each container runs an instance of a \models{Microservice}. We model a microservice as a composition of a \models{Runtime}, which provides the language-specific environment of the microservice, e.g. the Java Runtime. Microservices can also be composed of a number of \models{Frameworks}, which offer reusable components that streamline the development process, e.g., Django or gRPC. Lastly, microservices are also composed of the \models{Application Code} written by the development team.

In order to capture observability data, a cloud-native application needs to be instrumented through so-called \models{InstrumentationPoints}. Instrumentation refers to the process of embedding code within a software component that enables it to collect and emit metrics, traces and logs. Some components of the cloud-native deployment stack are equipped with so-called automatic instrumentation, meaning that instrumentation points are provided out-of-the-box. Depending on the observability data they collect, \models{InstrumentationPoints} can be of type \models{MetricInstrumentation}, \models{LogInstrumentation} and \models{TraceInstrumentation}. Each instrumentation type supports a different observability practice. 
While logging is similarly important for building reliable software systems, we mainly focus on modeling monitoring and tracing in this, as logging is less standardized to make a generic model beyond timestamped, human-readable text messages.

\textbf{Monitoring} is the continuous measurement of quantitative properties of a system over a period of time. \texttt{MetricInstrumentation}-Points emit \texttt{MetricEvents} which are timestamped value measurements. A \texttt{Metric}, in turn, represents the time-series collection of several \texttt{MetricEvents}. They may be simple \texttt{BaseMetrics}, collected directly through automatic instrumentation points provided in \texttt{Clusters}, \texttt{VMs}, \texttt{Containers}, \texttt{Runtimes} and \texttt{Frameworks}. In Figure \ref{fig:model} we show several of these types of \texttt{BaseMetrics}, but the list is not intended to be exhaustive. \texttt{Metrics} may also be specified and instrumented by developers in the case of \texttt{CustomMetrics}. While designing the monitoring of an application, practitioners face several design decisions, including (1) what \texttt{BaseMetrics} to collect from the available automatic \texttt{InstrumentationPoints}, (2) whether to add \texttt{CustomMetrics} and if yes, where to add the \texttt{InstrumentationPoints}, (3) how to configure the attributes \texttt{samplingRate} and \texttt{aggregationInterval} for each \texttt{Metric} instance.

\textbf{Tracing} enables end-to-end observation of appli\-cation behaviour by retrieving
and aggregating event data in so-called traces. In this context, a \models{TracingInstrumentation}-Point, which can be of type \models{AutomaticInstrumentation} if supported natively by the microservice \models{Runtime} or \models{Framework}, or of type \models{CustomInstrumentation} if added to the \models{ApplicationCode} by the development team, emits a \models{TraceEvent}. A pair of \models{TraceEvents} demarcating entry and exit build a \models{Span}, and multiple spans form a \models{Trace}. The information inside traces can further be processed into \models{TraceBasedMetrics}, e.g. the duration of spans or of entire traces. While designing the tracing of an application, practitioners again face several design decisions, namely (1) where to add \models{CustomInstrumentation} and (2) how to configure the attributes \models{samplingStrategy} and \models{samplingRate}.

For \textbf{Fault Detection}, observability systems employ so-called \models{FaultDetectionMechanisms} to be able to detect \models{Faults}. These mechanisms can employ a variety of detection methods, including pre-trained \models{Classifiers}, \models{Alerts},  or by having an trained site-reliability engineers look at \models{Dashboards}. For fault detection, practitioners first (1) need to select a feasible \models{FaultDetectionMechanism}, then (2) set appropriate attributes to not overload reliability teams and yet provide fast response to potential failures. Naturally, the prior design decisions on logging, monitoring and tracing strongly influence the quality of these detection methods, as too few inputs may lead to a low sensitivity, and too many observations lead to too much noise and thus may lead to oversensitivity.

\section{Approach to quantify observability effectiveness}\label{sec:concept}In order to arrive at informed and assessable observability design decisions, we need to be able to quantify their effectiveness. Observability serves as a necessary precursor for measuring many other system qualities, such as performance \cite{Ahmed_EffectivenessOfAPM_2016} or cost \cite{Kuhlenkamp_Costradamus_2017}, but observability itself is very challenging to quantify~\cite{Tamburri_MVCObservability_2018}. 
Its effectiveness is intricately connected with the very aspects it aims to capture. 
In this work, we look at observability for the specific purpose of reliability assurance, i.e., the process of ensuring that the system is running despite failures occurring.

Reliability assurance centers around two key metrics: (1) mean-time-to-failure / mean-time-between-failures ($MTTF / MTBF$) and (2) mean-time-to-restore ($MTTR$). $MTTR$ can be further divided into (2.1) mean-time-to-detect ($MTTD$) and (2.2) mean-time-to-repair ($MTTRepair$). 
The goal of maintaining a high $MTTF$ mostly falls beyond the domain of observability and instead relies on thorough testing. $MTTR$, conversely, is influenced by the presence of observability measures. 

Still, $MTTD$ and $MTTR$-like metrics aren't specific to observability mechanisms but apply to the entire operation process, including any other fault-tolerance mechanisms implemented. They serve to track the long-term effect of reliability measures and of reliability teams but are too coarse and can be influenced by many different overlapping factors.
For example, they do not indicate whether a failure could have been detected earlier or at less cost with a different observability configuration. To arrive at a systematic method for observability design decisions, we need to measure the impact of configuration and instrumentation decisions. Hence, we need to broaden our perspective beyond time-based process metrics. In the following, we propose a first set of fine-grained metrics as a basis of such a systematic method. 

\subsection{Concept: Fault Observability Metrics}\label{sec:metrics}
\begin{figure}[t]
    \centering
    \includegraphics[width=0.9\columnwidth]{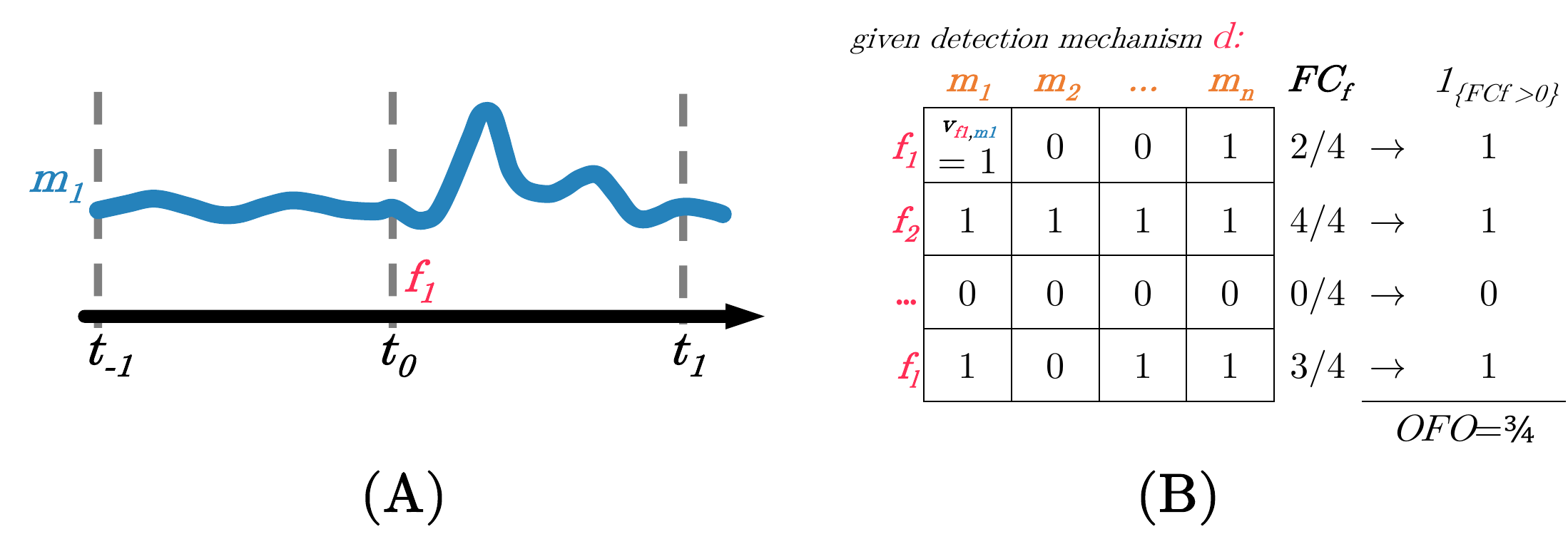}
    \caption{(A) Visualization of the fault model, used to define the metrics in (B)}
    \label{fig:faultmodel}
\end{figure}

To arrive at testable and explicit metrics for observability design decisions, we restrict our scope to the visibility of faults, or what we call \textbf{fault visibility}\cite{Borges_TracingServerless_2021}. Fault visibility can be defined as the degree to which the data produced during the occurrence of a specific fault is distinct enough from normal operation to trigger a fault detection mechanism. 
Intuitively, faults that produce very distinct data are easier and faster to troubleshoot. 
For \textbf{fault visibility}, we consider a set of faults $ F = \{f_1, f_2, ..., f_l\}$, a set of observability metrics or traces $M = \{m_{1},m_{2},...,m_{n}\}$ and a detection mechanism $d$. A fault $f$ can be considered visible in metric $m$ if the data recorded during the occurrence of the fault, represented by the time interval $[t_{0} - t_{1}]$, is significantly different from the data recorded under normal conditions, in time interval $[t_{\text{-}1} - t_{0}[$, and thus detected by detection mechanism $d$. Refer to \Cref{fig:faultmodel}(A) for a visual representation of these time intervals. 
As modeled in section \ref{sec:model1}, the detection mechanism can be of different types, e.g., a pre-trained classifier. We consider fault $f$ visible in metric $m$ if detection method $d$ is able to meet its detection threshold and detect the fault. If the detection mechanism is not able to detect the fault, it suggests that either (i) metric $m$ is unsuitable to detect $f$ (ii) the parameters of metric $m$ are not set appropriately or (iii) the function parameters of the detection mechanism $d$ are not tuned correctly. Fault visibility for fault $f$ in metric $m$ through detection mechanism $d$ can thus be defined as
\begin{equation}
    v_{f,m,d}=\left\{
    \setlength{\arraycolsep}{0.2cm}
    \begin{matrix} 
     1 & \quad  DF(m_{t}) > \alpha \\  
     0 & \quad \text{otherwise}  \\ 
    \end{matrix}\right.
\end{equation}
where $DF$ represents the detection function of $d$, $\alpha$ the configured threshold and $m_t$ a subset of observations of m for time $t$.

The visibility score can be determined for each fault and metric pair, allowing us to assess the impact of individual faults on specific metrics given their current configuration. 
However, these individual scores are not able to provide a comprehensive and generalized understanding of the observability of the system as a whole. To accomplish this, we can construct aggregate or composite scores. We propose two composite scores: fault coverage and overall fault observability. 
\textbf{Fault coverage} shows the degree to which a fault $f$ is visible across the set of different collected metrics $M$. We define it as the ratio between the number of collected metrics and a number of metrics where fault $f$ is visible. If a fault is not visible in any metric of the system, it implies that this fault is not covered by the system's observability, and thus the fault coverage for fault $f$ is 0.
\begin{equation}
    FC_{f,d}=\frac{1}{n}\sum_{i = 1}^{n} v_{f,m_{i},d}
\end{equation}

Lastly, the \textbf{overall fault observability} shows the ratio between the theoretically observable faults versus the faults that were actually observed or detected by detection mechanism $d$. This ratio can improve over time, as developers add more metrics to the instrumentation, or tune the configuration. It can then be defined as
\begin{equation}
    OFO_{d}= \frac{1}{l}\sum_{i = 1}^{l} 1_{\{FC_i > 0\}}
\end{equation}

Figure \ref{fig:faultmodel}(B) illustrates our suggested fault observability metrics using an example. The main goal for developers should be to increase the $OFO$. To achieve this, developers have three options, either increase the visibility of invisible faults by changing the configuration of existing metrics, add additional metrics that are more sensitive to the invisible faults, or make changes to the detection mechanism. 
To weigh between different options, developers need to consider the trade-off between observability overhead and cost. Thus, as a last metric for observability effectiveness, we propose a cost metric. Here, related work \cite{Ernst_OfflineTraceGeneration_2021,Reichelt_OverheadOpenTelemetryEtc,Dinga_EnergyEfficiencyOfMonitoring_2023} has already provided different methods and approaches to quantify the overhead of observability systems. For the purposes of this paper, we will use CPU utilization as a measure for cost, but other metrics such as memory and storage overhead are also suitable.

\subsection{Approach: Observability Experiments}\label{sec:experiments}
We propose an empirical and systematic approach to navigate the observability design space: observability experiments. %
This is a formalized process of experimenting inspired by Chaos Engineering. In Chaos Engineering, faults are injected into a running microservice system to test the system's ability to withstand faults. In our approach, rather than focusing on whether the system survives a fault, we are particularly interested in analyzing the %
quality of observability data generated during such fault experiments. This quality can be quantified with the metrics proposed in the previous section, and can be measured across various observability configurations and instrumentation alternatives. 
Practitioners can use observability experiments as a feedback loop to assess observability design decisions, specifically evaluating which changes in the instrumentation machinery and configuration improve or degrade the metrics, all the while keeping an eye on the cost of these changes.

In the following, we describe the approach briefly. An observability experiment formally consists of a description of the system under experiment (SUE), a workload, a set of treatments and a nonempty set of response variables.

\textbf{System Under Experiment (SUE)} In the context of observability experiments, the SUE is either the entirety of a system or a subset of it, i.e., only a select number of microservices. It is often unnecessary to deploy the entire microservice architecture when developers are only interested in studying the observability and instrumentation of a single service and its dependent services. Therefore, it is important to enable experimentation on subsets, especially considering that modern microservice architectures can quickly grow to hundreds of services \cite{Li_ObservabilityIndustrialSurvey_2022}.

\textbf{Workload} We require load generation to simulate users that create realistic observability data. Because the load can affect the data points of the observability data, we include it as a component of the experiment. Further, as a lot of the metrics devised in the previous section rely on comparative analysis, it is important to also have a way to generate a comparable load repeatedly.

\textbf{Treatments} are controlled changes to the system under experiment. 
Our conceptualization is broader than in Chaos Engineering, where treatment is chaotic or destructive. This need not be the case with treatments in observability experiments. We distinguish between fault treatments and instrumentation treatments. Instrumentation treatments allow users to easily change observability configuration without having to tinker with the SUE code and are executed during compilation. Fault treatments, in turn, are applied at runtime and change the SUE by means of dynamic fault injection.
Treatments allow us to investigate changes to the system typically beyond the scope of Chaos Engineering. For instance, the experiment operator might be interested in investigating if an increase in the sampling interval at some metric instrumentation point results in an increase in accuracy for a fault detection model. %

\textbf{Response Variables} in observability experiments are metrics or distributed traces and represent the different types of observability data that a system can emit. They are used to calculate the fault visibility, fault coverage and overall fault observability. Responses are decoupled from treatments. This means that we can define a response variable that is not directly related to a treatment, i.e., we could observe a response at service A while treating service B. The decoupling allows us to take into consideration higher-order effects of treatments. For example, we might wish to investigate whether injecting a network delay into service A consequently also affects the throughput at other services that depend on A. 

Such experiments are especially feasible in staging environments for systems where both the SUE and the observability system are described using infrastructure as code. In the following, we present a system to automate the process of observability assessments for microservice-based applications.

\section{OXN: Observability Experiment Engine}
\begin{figure}[b]
    \centering
    \includegraphics[width=\columnwidth]{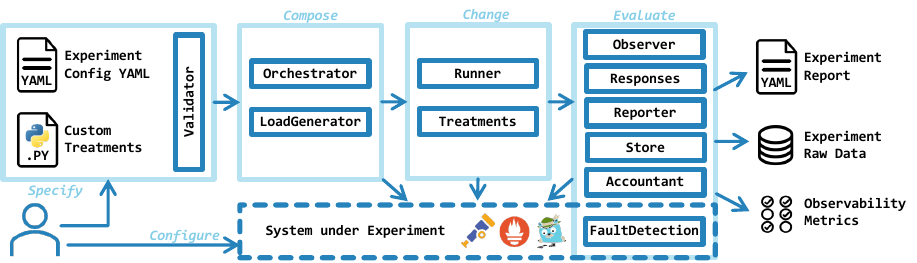}
    \caption{System architecture of \name{}}
    \label{fig:oxn-architecture}
\end{figure}

In this section, we present \name{}, an extensible software framework to run observability experiments\footnote{\url{https://github.com/nymphbox/oxn}}. 
\name{} follows the design principles for cloud benchmarking \cite{Silva_CloudBench_2013,2017-Bermbach-Book-CloudServiceBenchmarking,Bermbach_Benchfoundry_2017} and thus particularly strives for portable, repeatable, and relevant experiments.
We built \name{} around a \textsc{yaml}-based configuration file that allows experiments to be shared, versioned and repeated. 
The experiment configuration describes the SUE, i.e., deployment, treatments, workload and response variables  to collect (see \cref{sec:experiments}). 
Using this single experiment file \name{} orchestrates the complete experiment (see~\Cref{fig:oxn-architecture}), by first building and deploying the SUE as well as the load-generator. We use an extendable \textit{runner} component to execute different treatments, such as killing a service container, in combination with the automatic collection of experiment response variables through an observer component. 
For that, we partly rely on the SUE to implement the \textsc{OpenTelemetry} standard, with tools from the CNCF stack\footnote{\url{https://landscape.cncf.io}}, namely \textsc{Jaeger} and \textsc{Prometheus}. 

\subsection{Architecture}
The design of \name{} is modular, decoupled, and extensible. 
At its core, \name{} combines an \textit{orchestrator} to manage the SUE, a \textit{runner} to enact treatments, a \textit{load generator} to stress the SUE and \textit{observers} to capture results.
These components all use existing, well-tested software, which we loosely coupled together to enable observability experiments. Thus, each can be replaced or extended to adapt to different SUE or treatments.
For example, we used \textsc{Docker Compose} for the orchestrator. However, we could have also used a \textsc{Kubernetes}-based orchestrator or one based on \textsc{CloudFormation}. 
For the runner, we rely on an \textsc{Python} interface that can be used to manipulate the SUE.
For the load generator, we use the \textsc{Locust} framework, and for the observers, we implemented interfaces to query \textsc{Jaeger} and \textsc{Prometheus} for now.

Fault and instrumentation \textit{treatments} are defined in a flexible treatment model. %
\Cref{table:fault-treatment-library} shows the fault and instrumentation treatments implemented in our prototype. In addition to these treatments, custom treatments can be added easily by implementing a new treatment class against our treatment interface. This allows
\name{} to be extended with arbitrary fault scenarios or with new instrumentation approaches. Thus, allowing practitioners to build up a large library of relevant faults and treatments to test with \name{}.
\begin{table}[t]
    \caption{Different treatments implemented in \textsc{oxn}} \label{table:instrumentation-treatment-library}\label{table:fault-treatment-library}
    \resizebox{\columnwidth}{!}{\newcounter{treatmentsFoodnoteCounter}
\setcounter{treatmentsFoodnoteCounter}{\value{footnote}} \centering
\newcommand{\rightIndent}{\hspace{0.2cm}}
\begin{tabular}{lp{0.6\columnwidth}l}
    Name & Purpose & Tool \\ \hline
    \multicolumn{3}{l}{\textbf{Fault Treatment:}} \\
    \rightIndent{}Pause & Simulates an unresponsive service by suspending all processes in a service & \textsc{Docker} \\
    \rightIndent{}Kill & Simulates a service crash & \textsc{Docker} \\
    \rightIndent{}NetworkDelay & Injects network delay on an interface & \textsc{tc\footnotemark} \\
    \rightIndent{}PacketLoss & Injects packet loss on an interface & \textsc{tc} \\
    \rightIndent{}PacketCorruption & Injects packet corruption on an interface & \textsc{tc} \\
    \rightIndent{}Stress & Simulates resource exhaustion by injecting stressors & \textsc{stress-ng\footnotemark} \\ \hline
    \multicolumn{3}{l}{\textbf{Instrumentation Treatment:}} \\
    \rightIndent{}MetricSamplingRate & Changes the sampling interval for metrics & \textsc{Collector} \\
    \rightIndent{}TracingSamplingStrategy & Samples traces based on a given Strategy & \textsc{Collector} \\
    \rightIndent{}TracingSamplingRate & Samples traces based on a given sampling rate & \textsc{Collector}
\end{tabular}

}
\vspace{-1em}
\end{table}
\stepcounter{treatmentsFoodnoteCounter}%
\footnotetext[\value{treatmentsFoodnoteCounter}]{\href{https://linux.die.net/man/8/tc}{Linux traffic control}}
\stepcounter{treatmentsFoodnoteCounter}%
\footnotetext[\value{treatmentsFoodnoteCounter}]{\href{https://wiki.ubuntu.com/Kernel/Reference/stress-ng}{Linux kernel load and stress testing tool}}

Response Variables are captured through multiple components.
An \textit{observer} component captures experiment information, e.g., execution start and end timestamps.
The \textit{responses} component takes in metric and trace data, tracking defined variables and also implements different labeling strategies, depending on the data type.
A \textit{store} component writes all captured information as a binary data format that is readable by most data analysis solutions. 
A \textit{reporter} component can generate a machine-readable experiment report to give engineers a quick overview of the effectiveness of the examined observability design.

To calculate the fault visibility metrics \name{} offers an extensible interface, enabling developers to seamlessly integrate their own \textit{fault detection} mechanisms. This could range from threshold-based alerting techniques to more advanced anomaly detection models.
By default, we include a logistic regression classifier with \name{}, which can be automatically trained to detect faults for a given accuracy threshold, making it readily available for developers to use out-of-the-box. 
However, developers and observability practitioners can also integrate their own fault detection mechanism, enabling them to test both their observability configuration and their fault detection mechanisms separately. 
In this way, \name{} can also be used to evaluate fault detection mechanisms using real observability data instead of synthetic traces commonly used so far by research. 

Lastly, the \textit{accountant} component contains functionality to estimate costs of the given observability configuration based on CPU usage. This component could again be extended to include other cost metrics, such as storage or memory overhead.

\label{sec:architecture}
\section{Applicability and Exemplary Observability Design Assessment}\begin{figure*}[ht!]
    \centering
\includegraphics[width=\textwidth]{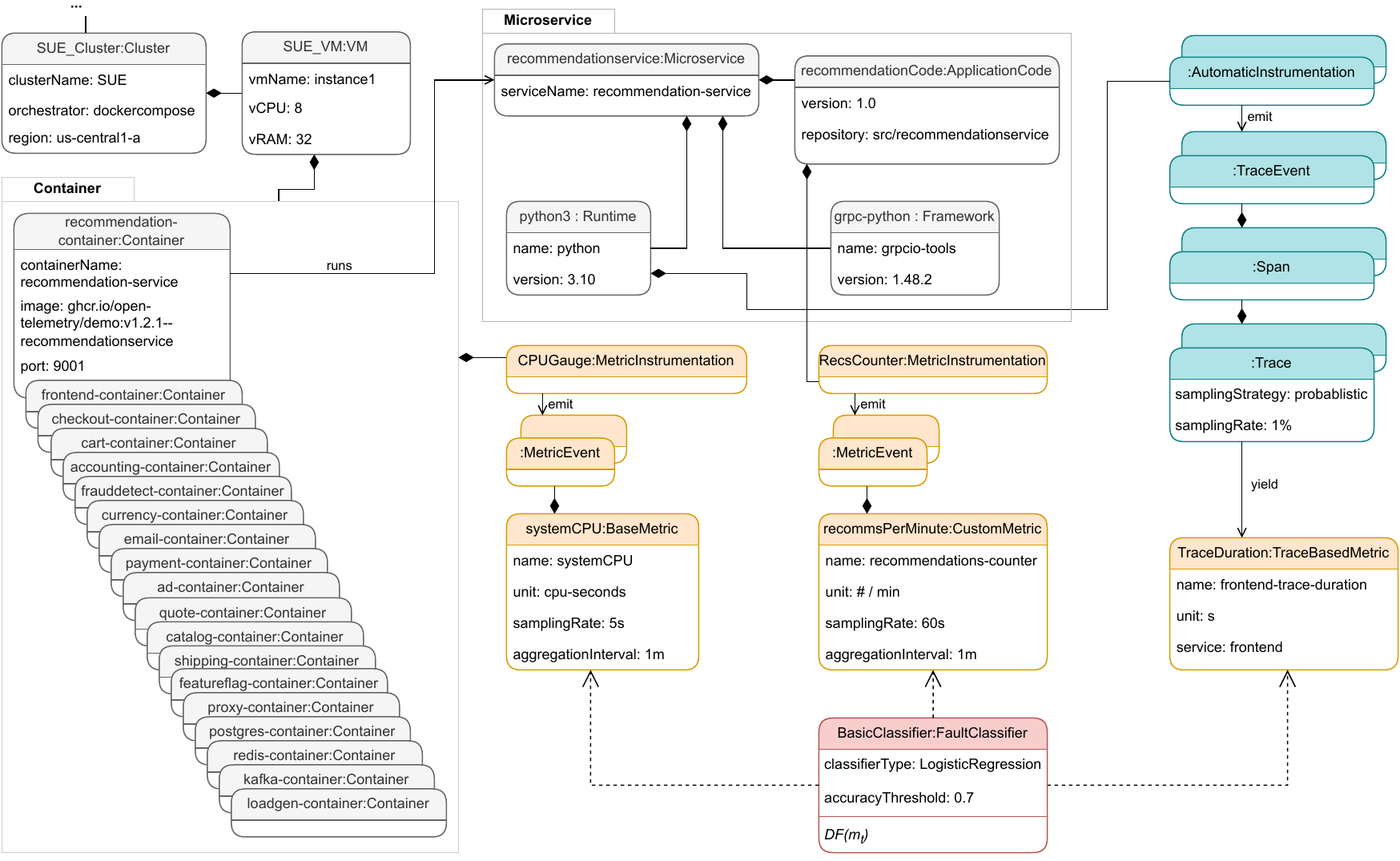}
    \caption{SUE with baseline observability configuration}
    \label{fig:sue}
\end{figure*}
Our approach and tooling allows practitioners to evaluate the observability of faults in their microservice application. Besides establishing the effectiveness of the current configuration, they can also use it to reason between observability design alternatives by weighing the observability-cost trade-offs. 
In this section, we demonstrate the applicability our approach by conducting an exemplary evaluation of the observability of a popular open source microservice application.

\subsection{SUE Setup}
As our system under experiment (SUE), we use the \textit{OpenTelemetry Astronomy Shop Demo}\footnote{\url{https://github.com/open-telemetry/opentelemetry-demo}} microservice application, which is a community project intended to illustrate the use of different observability methods and tools in a near real-world environment. The application consists of 20 core application services plus four dedicated for observability services. We chose this application because it covers a wide range of languages and frameworks across the cloud-native application stack. It is instrumented to collect traces across the whole application, leveraging both automatic and custom instrumentation. Besides traces, it is also instrumented to collect several metrics, including base container health metrics and custom metrics added manually by developers. 
It is a very suitable SUE to showcase OXN, since it offers such a broad spectrum of observability tuning knobs.   

For the experiments, we deploy a fork of the application\footnote{See the \name{} repo for details.} to ensure compatibility with the fault injection functionality of \name{} and to enhance container runtime monitoring. We run \name{} against the SUE on a cloud-based virtual machine (8vCPUs, 32GB Memory). Figure \ref{fig:sue} applies our model for observability design decisions (see \ref{sec:model1}) and shows the snapshot of our SUE at the time of the baseline measurement. We target our experiments on the recommendation service, which mirrors a real-life scenario for developers who, after creating a new service for their application, are faced with instrumentation and configuration decisions to be able to observe this service effectively.

Our baseline configuration monitors the overall CPU utilization of the system (\texttt{systemCPU}) by collecting and aggregating measurements of the \texttt{CPUGauges} of every container. By default, base metrics are configured with a 5s sampling rate.  Further, the baseline also collects the custom metric \texttt{recommsPerMinute}, through an instrumentation point that the developers added to the recommendation application code. This is configured with a 60s sampling rate by default. For tracing, the application is instrumented with both automatic and custom instrumentation, however the recommendation service in question leverages only automatic instrumentation. The tracing sampling strategy is set to a probabilistic sampler\footnote{\url{https://opentelemetry.io/docs/specs/otel/trace/tracestate-probability-sampling/}} with a 1\% sampling rate.

As a fault detection mechanism, we use the logistic regression classifier that is shipped with \name{} out-of-the-box. Logistic regression is conceptually simple, interpretable,  fast to train on large datasets and
only requires tuning one hyperparameter. %
For training, we split
the experiment data into training and test sets and ensure an equal class balance of fault and non-fault labels via an oversampling technique. %
We further preprocess the telemetry data by z-score normalization. %
After training, we evaluate the classifier's detection function (\textit{\texttt{DF(m\textsubscript{t})}}) on the test data and compute classification accuracy 
with a threshold of $0.7$  as our implementation of $\alpha$ for fault visibility. 

\subsection{Evaluating the Baseline - Results}
We investigate the observability of our SUE with three different types of faults, \texttt{Pause, PacketLoss} and \texttt{NetworkDelay}. For each fault, we run a 10min experiment under constant load (50 concurrent users) and repeat each experiment 10 times.

\begin{figure*}[t]
    \centering
    \begin{minipage}{0.52\textwidth}
        \centering
        \includegraphics[width=\linewidth]{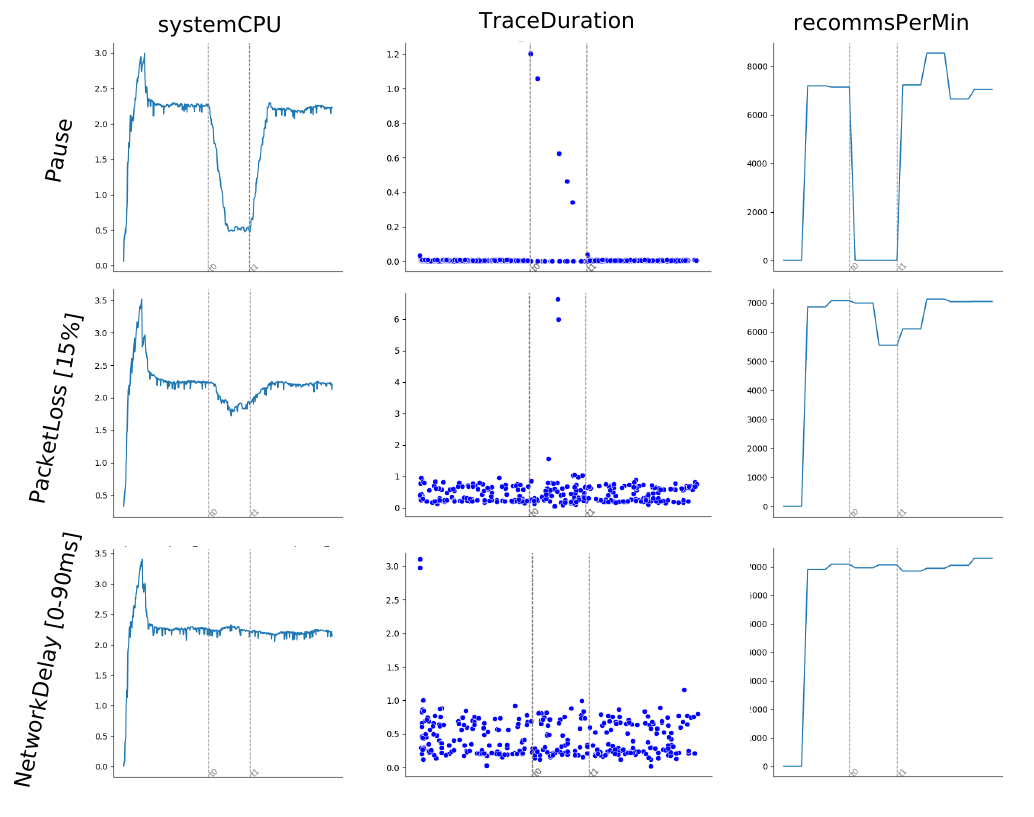}
        \vspace{1em}
        \caption{Experiments run against the SUE using \name{}, showing how different faults appear visually, similar to how a developer would see them in a dashboard. Note how the Pause fault is visible in all metrics. PacketLoss is noticeable in systemCPU but less pronounced in other metrics, with NetworkDelay not being visible at all. In \Cref{fig:plots2}, we see how the changes to the observability configuration proposed in \Cref{fig:alternatives} affect these metrics. }
        \label{fig:plots}
    \end{minipage}%
    \hfill
    \begin{minipage}{0.45\textwidth}
        \centering
        \includegraphics[width=\linewidth]{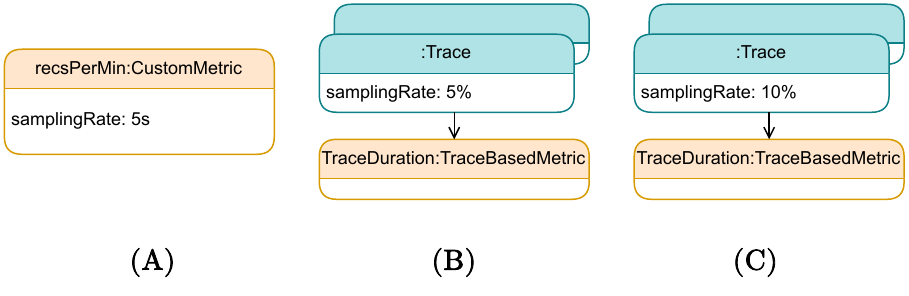}
        \caption{Observability design alternatives under consideration}
        \label{fig:alternatives}
        \vspace{1em}
        \includegraphics[width=\linewidth]{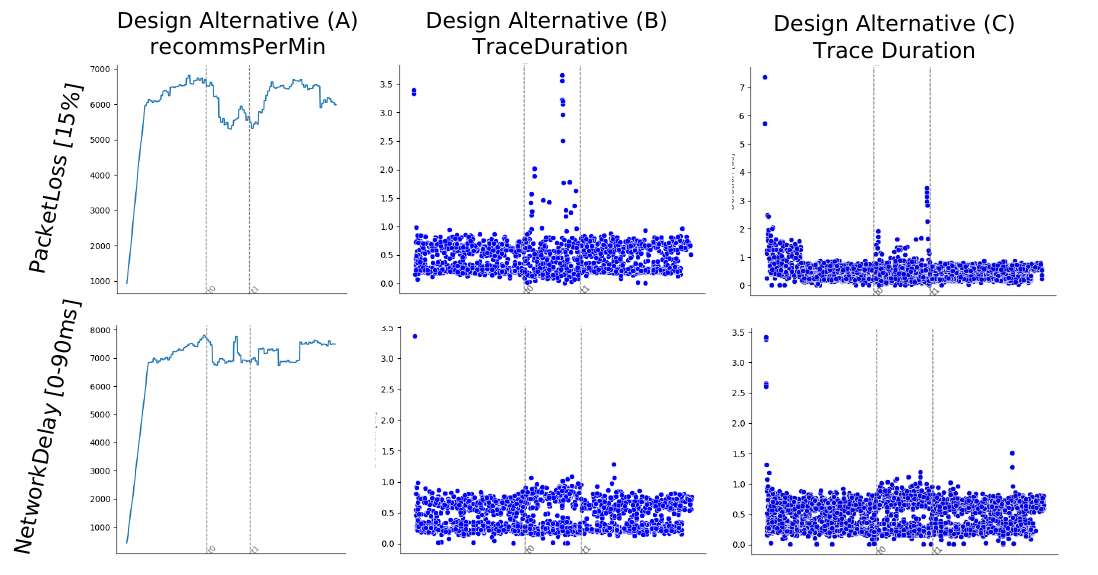}
       \caption{Visual changes to fault visibility across design alternatives. Recognize the now more pronounced effect of PacketLoss on \textit{recommsPerMin}, with NetworkDelay still not affecting this metric significantly. NetworkDelay causes a slight noticeable increase in TraceDuration for the longer Traces for both design alternatives B and C.}
        \label{fig:plots2}
    \end{minipage}
\end{figure*}

Figure \ref{fig:plots} plots the metrics we collected over the different experiments with our baseline observability configuration. %
\Cref{tabl:visibilityscores} shows the accuracy of the classifier, averaged over the ten experiment runs, and the resulting fault visibility scores. Because we set our classifier threshold $\alpha$ to 0.7, everything that falls below this threshold will not be identified as fault by our fault detection mechanism, and is therefore invisible.
As we can see, faults like the pause treatment, which simulates an unresponsive service, 
manifest themselves quite prominently across metrics. 
Packet loss is also present in the baseline observability configuration, though with a lower fault coverage (FC) score, since it only manifests in \texttt{systemCPU}. Lastly, faults like delay, which could, for example, simulate a faulty cache, are barely visible in the plots and are not detected by our fault detection mechanism, since the detection function doesn't reach the threshold for any of the collected metrics.

\begin{table}[h!]
\caption{Fault visibility metrics for default observability configuration} 
\centering
\resizebox{\columnwidth}{!}{
\begin{tabular}{l|lccl}
                      & CPU Util.           & Trace Dur.    & \#Recs/min           & FC         \\ 
\hline
Pause                 & $1 \quad (DF=0.83) $ & $1 \quad (1.00)$ & $1 \quad (0.89)$        & $3/3$      \\
PacketLoss [15\%]     & $1\quad (0.86)$       & $0\quad (0.61)$ & $0\quad (0.42)$        & $1/3$      \\
NetworkDelay [0-90ms] & $0\quad (0.50)$       & $0\quad (0.61)$ & $0\quad (0.43)$        & $0/3$      \\ 
\cline{5-5}
\multicolumn{1}{l}{}  &                     &               & \multicolumn{1}{l}{} & $OFO=2/3$ 
\end{tabular}
}
\label{tabl:visibilityscores}
\end{table}

\subsection{Evaluating Design Alternatives - Results}
To improve the fault observability of our application, we consider three observability design alternatives (see \Cref{fig:alternatives}). The first (Fig. \ref{fig:alternatives}A) is to increase the sampling rate of the application-level recommendation counter, which by default is set to only report once every minute. The other options under consideration are to increase the tracing sampling rate to either 5\% (Fig. \ref{fig:alternatives}B) or 10\% (Fig. \ref{fig:alternatives}C). 

For each observability design alternative, we repeated the  \texttt{PacketLoss} and \texttt{NetworkDelay} experiments, keeping everything else about the SUE and experiment setup the same. 
Figure \ref{fig:plots2} plots the collected metrics after the changes outlined above. \Cref{tab:redesign_results} shows the results regarding fault observability improvements while table  \ref{tabl:configchanges} reveals the costs associated with carrying out each of these observability design decisions (as CPU time [s] summed across the affected services). 

From these results, we can conclude that increasing the interval for the recommendation counter provides better fault coverage for \texttt{PacketLoss} faults, an increase from $1/3$ to $2/3$, but does not affect the $OFO$, since the \texttt{NetworkDelay} fault is still not visible. The second option provides better fault coverage for \texttt{NetworkDelay} faults, an increase from $0/3$ to $1/3$, as well as an increase to the $OFO = 3/3$, now that the delay fault is finally visible. The third option performs similar to the second one in the fault observability metrics. When looking at observability-related trade-offs in the form of cost (see also table \ref{tabl:configchanges}), we see that while B are C viable options to improve $OFO$, B is tied to a lower cost than C, with 3.05\%  overhead instead of 5.33\%. 

Thus, by applying the model, metrics and tooling presented in the paper, we can find a design decision with better fault observability, and we can immediately get a first performance impact assessment. A decision maker can now judge if the higher cost is acceptable and implement the change. Otherwise, more observability experiments can be performed to find a better solution.
Besides that, the model also provides developers and practitioners with a common language to understand, discuss and document their observability design decisions.

\begin{table}[]
    \centering
    \caption{Effects of the Design Alternatives on the fault observability metrics}
    \label{tab:redesign_results}
    \resizebox{\columnwidth}{!}{
        \begin{tabular}{c|ll|cc}
          & PacketLoss [15\%] & NetworkDelay [0-90ms] & $\Delta FC$ & $\Delta OFO$ \\ \hline
        (A) & 1 (DF=0.72)          & 0 (0.59)              & +1          & 0            \\
        (B) & 0 (0.58)          & 1 (0.77)              & +1          & +1           \\
        (C) & 0 (0.60)          & 1 (0.70)               & +1          & +1          
        \end{tabular}
        
    }
    \vspace{1em}
    \caption{Cost of the Design Alternatives in CPU time [s]} \label{tabl:configchanges}
    \centering
    \resizebox{\columnwidth}{!}{
    \begin{tabular}{l|rrrr}
     & Baseline SUE     & (A) & (B) & (C) \\ 
    \hline
    recomm.-service      & 143.30            & 150.74     & 143.86          & 144.63            \\
    otel-col             & 37.38             & 37.49      & 39.05           & 39.99             \\
    prometheus           & 9.01              & 9.07       & 9.07            & 9.48              \\
    jaeger               & 2.14              & 2.17       & 5.70            & 7.96              \\
    \hline
    total                & 191.83 & 199.47     & 197.68          & 202.06            \\ 
    
    overhead             & ~-                & +3.98\%     & +3.05\%          & +5.33\%           
    \end{tabular}
    }
\end{table}

\label{sec:eval}
\section{Limitations and Future Work}\label{sec:discussion}
While this paper represents an initial step in the development of observability metrics and the exploration of optimization strategies, it is important to acknowledge certain limitations. 
First, the approach relies on simulation and isolated experiments, which may not fully capture the complexities and faults faced in real-world applications. For this purpose, we deliberately designed \name{} with extensibility in mind, allowing developers to integrate their own load curves and custom faults. In the future, we aim to further validate the approach under real loads and more extensive fault scenarios.  

Second, the individual metrics proposed for fault observability were deliberately kept clear and flexible, to enable a technology-independent assessment. A generalization to more complex fault detection systems might not be possible without high coupling to the reporting mechanism.  Still, we consider our approach validated as we were able to show the impact of different configuration options on the generated observability data, evident both in the graphical plots and in the classifier score. As part of our ongoing effort to expand observability assessments, we aim to enhance our tool by incorporating support for alerting systems and Mean Time To Detect (MTTD). Besides this, we also aim to improve trade-off analysis by incorporating other cost metrics beyond CPU utilization.

Third, the current implementation of \name{} has some limitations, notably the absence of certain features like support for logs\footnote{At the time of implementation, \textsc{OpenTelemetry}  had not finalized logs.}, \textsc{Kubernetes} integration, among others. We aim to address these as we continue to utilize our tool in further observability assessments.
In future work, we plan to explore how observability decisions can be optimized and automated. Taking our metrics as a foundation, our next objective is to delve into intelligent and learning approaches to optimize and tune observability configuration parameters.

\section{Conclusion}As observability becomes an increasingly indispensable property of microservice applications, and configuration becomes increasingly complex, there is also a growing concern regarding how to best configure and instrument an application. To enable developers and practitioners to assess the suitability of different designs, we set out to make observability a testable and quantifiable system property, similar to software quality measures like test coverage.  

In this paper, we presented an approach for assessing and improving the fault observability of cloud-based microservice applications. Our approach consists of a model for understanding and documenting the observability design space, a set of metrics to assess fault observability, and a tool that provides quantitative evidence for observability design decisions beyond gut-feeling or professional intuition.
We showed how this approach can be used in practice, with an application of the model to document the observability design space of a state-of-the-art cloud-based microservice application. Further, we evaluated multiple designs, revealing the until now unexplored observability-cost trade-offs.
We aim to improve \name{} by adding support for more platforms such as Kubernetes, by integrating the tool in CI pipelines, and by increasing the automation and intelligence of the assessment process.
With the help of other practitioners, the model and tooling can also be extended to cover more fault scenarios and more diverse microservice environments. With this work, we lay the foundation toward a systematic method for supporting observability design decisions in cloud-native microservice applications.

\label{sec:conclusion}
\vspace{1em}
\section*{Acknowledgements}

\small \noindent Funded by the European Union (TEADAL, 101070186). Views and opinions expressed are however those of the author(s) only and do not necessarily reflect those of the European Union. Neither the European Union nor the granting authority can be held responsible for them.
\bibliography{refs}

\end{document}